\documentclass[]{tufte-handout}

\usepackage{amsmath}
\usepackage{booktabs} 
\usepackage{nicefrac}
\usepackage{units}
\usepackage{graphicx}
\usepackage{subfigure}
\usepackage{mathrsfs}

\title{A Tutorial on Variational Integrators}
\author{Stephen D. Webb}

\begin{document}
\maketitle

\section{Introduction}
\newthought{Symplectic integrators have been a staple} of accelerator physics for thirty years or more~\cite{ruth:83, ruth_forest:89, yoshida:90, forest_bengtsson_reusch:91, forest:06}. Their virtue is summarized quite nicely by Richard Talman as being ``exact tracking in an approximate lattice''. In this way, symplectic integrators are exactly Hamiltonian flows for Hamiltonians which approximate the one being integrated. The virtue of this is easy to see.

Consider the simple nonlinear Hamiltonian
\begin{equation}
\mathcal{H} = \frac{1}{2} p^2 + \frac{1}{2} q^2 - \frac{\lambda}{4} q^4
\end{equation}
for $q$ inside the bounded orbits. These orbits have to be bounded and periodic because this system is exactly integrable. Thus, there should be no spurious growth or decay of the Hamiltonian, which is a conserved quantity in the exact problem. This stability is unconditional on the time step, particularly for this one-dimensional example. Of course, more complex systems with separatrices will see these perturbations as causing dramatic changes in the underlying dynamics. But the key point is that symplectic integrators for single-particle dynamics are completely stable.

To see this, let us look at an example with a rather ambitious time step. For definiteness, we take $\lambda = 0.1$, and a time step $t = 0.3$. We take $6,000$ time steps, so this is approximately a few hundred periods.
\begin{marginfigure}
\includegraphics[scale=0.35]{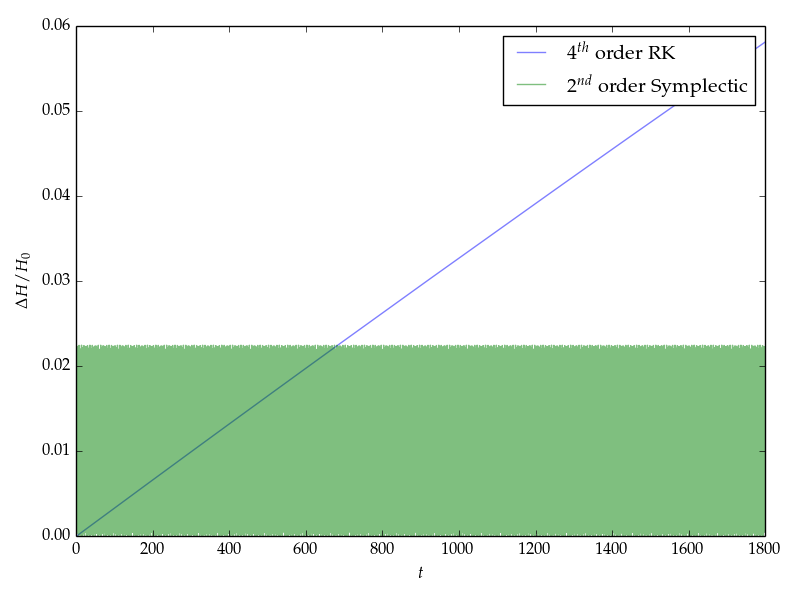}
\caption{$4^{th}$ order Runge-Kutta versus $2^{nd}$ order Symplectic integrator for conserving energy over many oscillations of the system.}
\end{marginfigure}
We note that, for short times, the $4^{th}$ order Runge-Kutta scheme is ``more accurate'' than the $2^{nd}$ order Symplectic integrator. However, the growth in energy from RK4 is monotonic, and the system is forever spiraling. This is unphysical for our energy-conserving bounding double-well potential. The S2 integrator, on the other hand, is always oscillating around the original energy. This illustrates the importance of symplectic integrators -- over many, many periods of a system, the symplectic condition keeps the numerical solutions close to the actual solutions.

The reason for this result is simple enough -- $4^{th}$ order Runge-Kutta does not care that the system of differential equations is derived from an action principle, while the symplectic integrator does. Specifically, the equations of motion come from the action
\begin{equation}
S = \int  p \dot{q} - \mathcal{H}(p, q) ~ dt
\end{equation}
One method of obtaining the symplectic integration scheme is through operator splitting (see the Yoshida paper and the Forest review). This has been the dominant approach for accelerator physics because of how simple it is to derive and implement. An alternative approach, first discussed in detail by Marsden and West~\cite{marsden_west:01}, is to study the problem by discretizing the action. Hence, we can obtain a second order symplectic integrator with the discretization
\begin{equation}
\begin{split}
S =\sum_n \biggl \{  \left [ \frac{p_{n+1/2} + p_n}{2} \left ( \frac{q_{n+\nicefrac{1}{2}} - q_n}{t/2} \right ) - \mathcal{H}(p_{n+1/2}, q_{n+1/2}) \right ] + \\ \left [ \frac{p_{n+1/2}+p_{n+1}}{2} \left ( \frac{q_{n+1} - q_{n+\nicefrac{1}{2}}}{t/2} \right ) - \mathcal{H}(p_{n+1/2}, q_{n+\nicefrac{1}{2}})\right ] \biggr \} t
\end{split}
\end{equation}
by minimizing the action using a discretized Euler-Lagrange set of equations. This has a number of advantages over the splitting method. One is that it naturally generates implicit integrators, and does not care that a Hamiltonian may not be splittable. The other is that it easily generalizes to action integrals on fields.

\section{Multisymplectic Integrators}

\newthought{Most field equations are derived from a Lagrangian} action principle, such as the electromagnetic Lagrangian
\begin{equation}
\mathcal{L}_{EM} = \frac{1}{8 \pi} F^{\mu \nu} F_{\mu \nu}
\end{equation}
Marsden \emph{et al.}~\cite{marsdenPatrickShkoller:98} discussed the \emph{multisymplectic} geometric structure of such systems. These systems conserve a two-form, and their flows are associated with conservation laws described by Noether's theorem. They also went on to derive discretized versions of these actions, which led to exactly conserved quantities and stable dynamics for the equations of motion.

Stamm, Shadwick, and Evstatiev~\cite{shadwick:13} took the next logical step, discretizing the entire Low Lagrangian~\cite{low:58},
\begin{equation}
\mathcal{L} = \int d\vec{x} d t \left (\left \{ \int  \left ( T(\vec{v}) + \frac{e}{c} \vec{v} \cdot \vec{A} - e \varphi \right ) f(\vec{x}, \vec{v}, t) d\vec{v} \right \} + \frac{1}{8 \pi} F^{\mu \nu} F_{\mu \nu} \right ),
\end{equation}
to obtain particle-in-cell algorithms. In this way, they have developed a method for creating multisymplectic plasma simulations. The advantage here is obvious from our preceding discussion -- conservation laws deeply constrain the motion, so that every multisymplectic particle-in-cell algorithm is "exact solution to an approximate plasma". In this way, simulations of plasmas preserve some discretized version of the conserved quantities from the continuum limit. This also leads to some insight into the origins of grid heating and charge conservation.

Charge conservation is a Noetherian current of the Low Lagrangian. On structured meshes, such as the Yee cell, this has a discrete symmetry corresponding to the particle shape and how it couples to the fields. This explains the difficulty of getting charge conservation on an unstructured mesh -- unstructured meshes have no symmetry group, and hence no Noetherian current!

Similarly, grid heating arises from the demand for momentum conservation in a system that no longer has continuous translational symmetry\footnote{This observation is due to Brad Shadwick, and is a fantastic insight into how we can break our simulations by insisting on conserved quantities that the simulation does not want to conserve.}. By relaxing this requirement, we can bypass grid heating by simply accepting the lack of momentum conservation \emph{per se}, but rather, for example, a crystal momentum type conservation law. 

\section{One-Dimensional Electrostatic PIC}

To illustrate the nuts and bolts of computing a multisymplectic particle-in-cell algorithm, we consider a one-dimensional electrostatic problem. This glosses over many of the problems still encountered in the electromagnetic limit, but does show the mechanics of obtaining these algorithms from a discretization. The Low Lagrangian for such a system is
\begin{equation}
\mathcal{L} = \int dx \left (\left \{ \int  \left ( \frac{1}{2} m_e \dot{x}^2 - e \varphi \right ) f(x, \dot{x}, t) dv \right \} + \frac{1}{8 \pi}(\partial_x\varphi)^2 \right )
\end{equation}
Now, we discretize. We choose for our basis function of $\varphi$ mound functions spanning one grid spacing, so that
\begin{equation}
\varphi = \sum_n \Psi(n, x)\varphi_n
\end{equation}
where
\begin{equation}
\Psi(n, x) = \biggl \{ \begin{array}{cc}
1 - \left |\frac{(x - n \Delta x)}{\Delta x}\right | & x \in \bigl( (n-1)\Delta x, (n+1) \Delta x \bigr]  \\
0 & \textrm {otherwise}
\end{array}
\end{equation}
These are the usual tent functions of finite element codes, but regularly arrayed in $x$ so that the grid is structured. This choice of function can exactly represent a linearly varying electric field. This is not necessary, though. Many other options are available -- one could use a parabola two cells wide for a higher order method, or to get smoothness use a parabolically damped Gaussian that vanishes smoothly at the cell edges, or sinc functions. For simplicity we consider the tent function.

Inserting this into the Lagrangian gives
\begin{equation}
\begin{split}
\mathcal{L} = \int dx ~ \int  \Bigg\{ \underbrace{\left ( \frac{1}{2} m_e \dot{x}^2\right ) f(x, \dot{x}, t)}_{\textrm{kinetic energy}} -\\
\underbrace{e \sum_n \Psi(n, x)\varphi_n   f(x, \dot{x}, t)}_{\textrm{charge deposition}} \Bigg \}d\dot{x} \\
 + \underbrace{\frac{1}{8 \pi}\left ( \sum_n \partial_x\Psi(n, x)\varphi_n \right)^2 }_{\textrm{field energy}}
 \end{split}
\end{equation}
The expansion of $\varphi$ into local basis functions gives the spatial variation, and the sums and integrals can be exactly solved to obtain a local stencil. For completeness, we evaluate the stencil for this system. This type of procedure will arise over and over in the derivation of these algorithms, and we therefore will walk through all the steps carefully.

Expanding the series, we get
\begin{equation}
\sum_m \sum_n \int dx \left (\partial_x\Psi(n, x) \partial_x\Psi(m, x) \right ) \varphi_m \varphi_n = \overline{\varphi}^\dagger \mathbb{S} \overline{\varphi}
\end{equation}
where $\overline{\varphi}$ is the array of potential coefficients, and $\mathbb{S}$ is the stencil from evaluating the integral
\begin{equation}
\mathbb{S}_{m, n} = \int dx \left (\partial_x\Psi(n, x) \partial_x\Psi(m, x) \right )
\end{equation}
Because the shape functions are only one cell wide, the matrix $\mathbb{M}$ is tridiagonal, and zero everywhere else. Because of our tent functions, this stencil very simply becomes
\begin{equation}
\mathbb{S} = \frac{1}{\Delta x}\left( \begin{matrix}
2 & -1 & & 0\\
-1 & \ddots & \ddots &  \\
  & \ddots & \ddots &  -1 \\
0 &  & -1 & 2  \end{matrix}  \right)
\end{equation}

We next expand the phase space density in a macroparticle basis, which is a moveable finite element basis with delta function velocities. There is nothing stopping us from choosing the basis function for the phase space density to be whatever we like, independent of the basis functions for the fields. Hence, we leave it arbitrary as:
\begin{equation}
f(x, \dot{x}, t) = \sum_i \Phi(x-x_i(t)) \delta(\dot{x} - \dot{x}_i(t)) W_i
\end{equation}
where $W_n$ is the macroparticle weight, $x_n$ its position, and $\dot{x}_n$ its velocity. Thus we have a parabola centered at $x_i(t)$ with width $\Delta x$ that vanishes at one cell length. Again, we are free to choose any basis we like -- the elegance of this formalism is that the field shape function and the particle shape functions are decoupled. This stands in contrast with the usual treatments of PIC~\cite{birdsallLangdon:85, hockneyEastwood:88}, where the field interpolation and charge deposition are carefully coupled to each other to prevent spurious forces. Where the normal treatments insist on imposing certain symmetries and dynamics (momentum conservation, energy conservation, \emph{etc.}), the discretized action approach generates analogs of these properties naturally.

Let us assume, for the sake of simplicity, that we have a real number of particles. In some beam dynamics simulations, this is becoming the norm thanks to the growth of supercomputer facilities and the improvement of codes. Thus, $\Phi(z) = \delta(z)$ is our basis function, with $W_i = 1$. This simplifies the convolution integral between $f$ and $\varphi$ dramatically, although it does require a real number of particles to be reasonable.

Carrying out the spatial and velocity integrals yields the spatial discretization of the Lagrangian:
\begin{equation}
\mathcal{L} =  \sum_i \frac{1}{2} m_e \dot{x}_i^2  + e \sum_i \sum_n \Psi(n, x_i) \varphi_n + \frac{1}{8 \pi} \overline{\varphi}^{\dagger} \mathbb{S} \overline{\varphi}
\end{equation}
To obtain the time derivatives multisymplectically, we can take a discretization in time as discussed by Marsden \& West. For simplicity, we consider a first order in time discretization -- by using the methods described in the \emph{Acta Numerica} paper we may obtain higher order integrators. Here, we take 
\begin{subequations}
\begin{equation}
\dot{x}_i = \frac{x_i^{(m+1)} - x_i^{(m)}}{t}
\end{equation}
\begin{equation}
 x_i = x_i^{(m)}
 \end{equation}
 \begin{equation}
 \overline{\varphi} = \overline{\varphi}^{(m)} 
 \end{equation}
 \end{subequations}
 where $m$ designates the discrete time and $t$ is the time step. It is important to note that in the Marsden \& West formulation, there is no $v$ and therefore no tangent vector to the configuration space. This is a subtle and important point -- the symplectic two-form \emph{does not} depend on the velocity for a discretized Lagrangian. We furthermore make an important convention note here: in the work by Marsden, the \emph{discretized Lagrangian} in his language has the units of action, i.e. it is given by $L dt$. Hence, our discretized Lagrangian is actually the approximate action.
 
 This gives the fully discretized Lagrangian as
 \begin{equation}
 \begin{split}
\mathbb{L}_D(m, m+1) = &\underbrace{\sum_i \frac{1}{2} m_e  \frac{ \left (x_i^{(m+1)} - x_i^{(m)} \right )^2}{t}}_{\textrm{particle move}} + \dots \\
& + \underbrace{e \sum_i \sum_n \Psi \left (n, x_i^{(m+1)} \right ) \varphi_n^{(m+1)} t}_{\textrm{deposition/interpolation}} + \underbrace{\frac{1}{8 \pi} \overline{\varphi}^{(m+1)\dagger} \mathbb{S} \overline{\varphi}^{(m+1)} t}_{\textrm{field solve}}
\end{split}
 \end{equation}
 The underbrace indicates what parts of a typical PIC algorithm each part of the discretized Lagrangian contains. We see that charge deposition and force interpolation are hopelessly entwined -- we already knew this from exercises in Hockney \& Eastwood illustrated the need for the force interpolation shape function to be the derivative of the charge deposition shape function. This result arises naturally here, not out of an appeal for avoiding spurious forces and carefully balancing them to zero, but from simply requesting our system obey symplecticity.
 
Applying the discrete Euler-Lagrange equations\footnote{A derivation of this may be found in the Marsden \& West paper, but it follows the same derivation procedure as for continuum variational principles.} to this Lagrangian yields the finite difference equations
\begin{subequations}
\begin{equation}
\begin{split}
m_e \left ( \frac{x_i^{(m+1)} - x_i^{(m)}}{t} - \frac{x_i^{(m+2)} - x_i^{(m+1)}}{t} \right )  = \\ - e \sum_n \partial_{x}\Psi \left (n, x_i^{(m+1)} \right ) \varphi_n^{(m+1)} t
\end{split}
\end{equation}
\begin{equation}
\frac{1}{4 \pi} \mathbb{S} \overline{\varphi}^{(m+1)} = e \sum_i \overline{\Psi} \left ( x_i^{(m+1)} \right )
\end{equation}
\end{subequations}
where $\overline{\Psi}$ is the array of shape functions with the same order as $\overline{\varphi}$. As becomes clear here, we can conveniently introduce the velocity as an intermediate variable
\begin{equation}
v^{(m)} = \frac{x^{(m+1)} - x^{(m)}}{t}
\end{equation}
but it does not explicitly appear in the discretized Lagrangian mechanics. We get the ``correct'' interpolation of forces combined with charge deposition for free. We also get the correct discrete Poisson equation. All this from the discretized action.

\section{Conclusion}

We have thus discussed the concept of discretized Lagrangians as an approach to obtain self-consistent particle-in-cell algorithms. This approach is desirable because it constrains the algorithm to obey the symplectic condition, which in 1D is equivalent to phase space conservation, but in higher dimensions is much more restrictive. This is quite an important step, as there is a great temptation to assume that discrete quantities should have exact analogs to continuum quantities. As the point about discretized space breaking momentum conservation should illustrate, discretized systems are in many ways fundamentally different from their continuum limits.

We have illustrated the nuts and bolts of deriving multisymplectic particle-in-cell algorithms with a one-dimensional electrostatic example. Many subtleties remain, such as gauge invariance in electromagnetics, charge conservation, and the more complex symmetry groups of the system.

\bibliography{variationalBib}

\begin{thebibliography}{11}
\providecommand{\natexlab}[1]{#1}
\providecommand{\url}[1]{\texttt{#1}}
\expandafter\ifx\csname urlstyle\endcsname\relax
  \providecommand{\doi}[1]{doi: #1}\else
  \providecommand{\doi}{doi: \begingroup \urlstyle{rm}\Url}\fi

\bibitem[Birdsall and Langdon(1985)]{birdsallLangdon:85}
C.~K. Birdsall and A.~B. Langdon.
\newblock \emph{Plasma Physics via Computer Simulation}.
\newblock McGraw-Hill, New York, 1985.

\bibitem[Forest(2006)]{forest:06}
\'{E}. Forest.
\newblock Geometric integration for particle accelerators.
\newblock \emph{J. Phys. A: Math. Gen.}, 39:\penalty0 5321--5377, 2006.

\bibitem[Forest and Ruth(1989)]{ruth_forest:89}
E.~Forest and R.~D. Ruth.
\newblock Fourth-order symplectic integration.
\newblock \emph{Physica D}, 43\penalty0 (1):\penalty0 105--117, 1989.

\bibitem[Forest et~al.(1991)Forest, Bengtsson, and
  Reusch]{forest_bengtsson_reusch:91}
E.~Forest, J.~Bengtsson, and M.F. Reusch.
\newblock {Application of the Yoshida-Ruth techniques to implicit integration
  and multi-map explicit integration}.
\newblock \emph{Phys. Lett. A}, 158\penalty0 (3):\penalty0 99--101, 1991.

\bibitem[Hockney and Eastwood(1988)]{hockneyEastwood:88}
R.~W. Hockney and J.~W. Eastwood.
\newblock \emph{Computer Simulation Using Particles}.
\newblock Institute of Physics Publishin, 1988.

\bibitem[Low(1958)]{low:58}
F.~E. Low.
\newblock {A Lagrangian Formulation of the Boltzmann-Vlasov Equation for
  Plasmas}.
\newblock \emph{Proc. R. Soc. London Ser. A. Math. Phys. Sci.}, 248\penalty0
  (1253):\penalty0 282--287, 1958.

\bibitem[Marsden and West(2001)]{marsden_west:01}
J.~E. Marsden and M.~West.
\newblock Discrete mechanics and variational integrators.
\newblock \emph{Acta Numerica}, 2001.

\bibitem[Marsden et~al.(1998)Marsden, Patrick, and
  Shkoller]{marsdenPatrickShkoller:98}
Jerrold~E. Marsden, George~W. Patrick, and Steve Shkoller.
\newblock Multisymplectic geometry, variational integrators, and nonlinear
  pdes.
\newblock \emph{Comm. Math. Phys.}, 199:\penalty0 351--395, 1998.

\bibitem[Ruth(1983)]{ruth:83}
R.~D. Ruth.
\newblock A canonical integration technique.
\newblock \emph{IEEE Trans. Nucl. Sci.}, NS-30\penalty0 (4), August 1983.

\bibitem[Stamm et~al.(2014)Stamm, Shadwick, and Evstatiev]{shadwick:13}
A.~Stamm, B.~Shadwick, and E.~Evstatiev.
\newblock {Variational Formulation of Macro-Particle Models for Electromagnetic
  Plasma Simulations}, 2014.
\newblock URL \url{arXiv:1310.0450v2 [physics.comp-ph]}.

\bibitem[Yoshida(1990)]{yoshida:90}
H.~Yoshida.
\newblock Construction of higher order symplectic integrators.
\newblock \emph{Phys. Lett. A}, 150\penalty0 (5-7):\penalty0 262--268, 1990.

\end{thebibliography}

\end{document}